# Quantifying Legibility of Indoor Spaces Using Deep Convolutional Neural Networks: Case Studies in Train Stations

Zhoutong Wang*[a, b], Qianhui Liang*[a,1], Fabio Duarte[a, c], Fan Zhang[a], Louis Charron[a], Lenna Johnsen[a], Bill Cai[a], Carlo Ratti[a]

* Equal Contribution
[a] Senseable City Laboratory, Massachusetts Institute of Technology, 77 Massachusetts Avenue, Cambridge, MA 02139, USA
[b] Graduate School of Design, Harvard University, 48 Quincy, St. Cambridge, MA 02138, US
[c] Pontificia Universidade Catolica do Parana, Rua Imaculada Conceicao, 1155 Curitiba, PR, Brazil
[1] Corresponding author: 1. email address: qianhuil@mit.edu; 2. postal address: 70 Pacific Street 739B, 02139, Cambridge, MA, USA

**Abstract**

*Legibility is the extent to which a space can be easily recognized. Evaluating legibility is particularly desirable in indoor spaces, since it has a large impact on human behavior and the efficiency of space utilization. However, indoor space legibility has only been studied through survey and trivial simulations and lacks reliable quantitative measurement. We utilized a Deep Convolutional Neural Network (DCNN), which is structurally similar to a human perception system, to model legibility in indoor spaces. To implement the modeling of legibility for any indoor spaces, we designed an end-to-end processing pipeline from indoor data retrieving to model training to spatial legibility analysis. Although the model performed very well (98% top-1 accuracy) overall, there are still discrepancies in accuracy among different spaces, reflecting legibility differences. To prove the validity of the pipeline, we deployed a survey on Amazon Mechanical Turk, collecting 4,015 samples. The human samples showed a similar behavior pattern and mechanism as the DCNN models. Further, we used model results to visually explain legibility in different architectural programs, building age, building style, visual clusterings of spaces and visual explanations for building age and architectural functions.*

**Keywords**:

Indoor space legibility; Deep Convolutional Neural Network; Human perceptions

## 1. Introduction

Legibility refers to the ability of spaces to be recognized and identified by their users from their visual qualities. As Kevin Lynch put it, legibility refers to "the ease at which its (spatial) parts may be recognized and can be organized into a coherent pattern" (Lynch, 1960, p2). Lynch proposed that legibility is the key to building a *cognitive map*, or an internal representation of an environment which individuals use as a reference when navigating a space. Therefore, the legibility of the space impacts people's ability to locate themselves, comprehend multiple elements as part of a larger spatial whole, and ultimately to find their way—or navigate space (Herzog et al, 2003).

Lynch's work focuses on legibility at the city level, while some other authors have also applied the concept as a measurement of navigation in architectural spaces (Passini, 1992). Legibility is particularly important in indoor spaces primarily used for commuting, such as train station and airports. In successful cases, commuters and visitors alike can easily recognize and identify the spaces around them, resulting in reduced navigation time increased understanding of the space. If one were able to quantify this quality, such spaces should reflect a higher level of legibility. Although the legibility of indoor spaces can be easily perceived by any individual, a reliable and robust method to measure legibility is absent from the current literature.

We utilized a Deep Convolutional Neural Network (DCNN) to model legibility through collected images in indoor spaces. To start with, we designed a parsimonious device through which we collected over 200,000 images in Gare de Lyon and Gare St. Lazare in Paris. Subsequently, we trained a DCNN to predict the location (defined by spatial segments) of every image. The model achieved 98% accuracy on testing set. Analysis of the model's results suggested some recommendations and guidelines for spaces.

To validate the correctness and robustness of the model, we deployed a survey on Amazon Mechanical Turk to confirm the consistency in the decision making process between DCNN and human performance. The result strongly implies similarity between the two.

## 2. Literature Review

### a. Difficulty in measuring legibility

Although most research discusses the universality of legibility ([Weisman, 1981; Herzog et al. 2003; Soltani et al. 2016](#)), other works suggest that legibility relies on individual experience, meaning that each person has a unique understanding that originates from his or her socio-cultural associations ([Gulick, 1963](#)) or personal living experiences ([Yadav, 1987; Duarte, 2017](#)).

However, experiments designed to measure individual experiences are labor intensive. Moreover, in spite of the importance of understanding different spatial perceptions, public spaces need to take into consideration the collective legibility. As a result, several scholars have explored different methods to measure the collective spatial legibility. During the last five decades, there have been two major methodologies to assess indoor legibility: qualitative study through questionnaires and behavior monitoring; and quantitative study using modelling and simulation.

### i. Qualitative analysis

Many surveys and participant observation methods have been proposed to measure legibility by asking participants about their perception and opinions about certain spaces. O'Neill (1991) assessed the influence of topological connections between points on cognitive mapping and wayfinding performance by surveying 63 participants within three building settings. Weisman (1980) used self-reported data to evaluate the legibility of 10 buildings. Notwithstanding, these observation and survey methods are expensive, time consuming, and raised concerns about individual biases ([Downs et al. 1977](#)). Therefore, these efforts have been limited in scope. Lynch's groundbreaking work, for instance, was criticized for using its small sample to generalize, using only around 30 interviews in Boston and 15 in Jersey City and Los Angeles, meaning the results may show highly personal biases ([Strohecker, 2000](#)).

### ii. Quantitive analysis

In recent years, researchers have utilized quantitative methods such using space syntax to measure legibility. A more recent approach is to use visual graph analysis (VGA) and agent-based analysis on space syntax representations to understand the relationship between legibility and different building typologies ([Soltani, et al. 2016](#)). However, the claim that space syntax methods accurately model pedestrian choice making has faced criticism ([Ratti, 2003](#)). Moreover, even if human behavior is correctly modeled, this approach only considers building typology and spatial configuration while ignoring other visual influential elements of the space.

With the development of mobile geographic information systems (GIS) and mobile mapping technologies, some researchers have proposed experiments using global positioning systems (GPS) to capture participants' movement and use that data to quantify legibility ([SA Al-ghamdi, 2015](#)). However, even if it is possible to overcome the technical difficulties associated with indoor localization, it remains arduous to integrate spatial attributes with the collected movement information.

### iii. Current challenges and limitations

Following the many efforts described above, we conclude that the study of legibility has encountered two major challenges:

First, it is challenging to integrate the influence of different spatial attributes in the evaluation process. Generally, legibility of space is considered to be influenced by three major factors: the floor-plan configuration, its complexity and the saliency of architectural components in the environment ([Hunt, 1984; Koseoglu et al. 2001; O'Neill, 1991](#)). Most of the previous works only taking into account one or two factors into account at most. Therefore, those experiments might omit some other factors outside of assumption and thus failed to reach a holistic conclusion. Even if it is feasible to take into account all possible factors, it is very difficult to decide the weights for every factor.

Second, it is very difficult to confirm the validity of human behavior and preferences in the measurement of legibility. In most qualitative methods, subjective perception is largely in question due to small sample sizes. Socio-cultural differences ([Gulick, 1963](#)), individual experiences ([Yadav, 1987](#)) may induce the biases in such an approach. In quantitative methods where behaviors of users are monitored to study legibility, researchers cannot provide strong evidence of cause and effect. The user behaviors may be triggered by events other than the legibility of space. To be further applied in architectural design and space management, a more objective method of legibility evaluation is needed in practice.

**b. Images, better representations**

We've all heard the cliché, "a picture tells a thousand words". Images not only contain the labels assigned to its pixels, but have also been embedded with a wide range of attributes from a collection of geometry and objects to their configuration ([Khosla, 2014](#)). Information such as objects ([Ren, et al. 2015](#)) and scenes ([Zhou et al. 2014](#)) can be inferred from images. Spatial knowledge such as depth ([Torralba, et al. 2002; Liu, et al. 2015](#)), 3D scene construction ([Koutsoudis, et al. 2015](#)), human perception ([Zhang et al. 2018](#)) and spatial configuration ([Peng, et al. 2017](#)) have also been explored in recent years using images. With so many features embedded in images, images are good representations of spatial attributes compared with the topological figure-ground information adopted by spatial syntax. Therefore, images are a rich source for a quantitative study of legibility.

**c. Deep Convolutional Neural Network**

A Deep Convolutional Neural Network (DCNN) is a popular architecture widely applied in interpreting images. DCNNs are very similar to human visual systems: both use restricted receptive fields and a hierarchy of layers which progressively extract more and more abstracted features ([Kheradpisheh, et al. 2016](#)). Studies have shown DCNNs to resemble human feed-forward vision in invariant object recognition ([Geirhos, et al. 2017; Majaj, et al. 2018; Schrimpf, et al. 2018](#)) and human-level concept learning can be done through probabilistic program induction ([Lake, et al. 2015](#)).

## 3. Methodology

To resolve the two key issues described above, we propose a pipeline that starts with the utilization of a portable and dismantable device to collect images from indoor spaces followed by the assessment of spatial legibility through a probabilistic inference by Deep Convolutional Neural Networks. For validation, we design a parallel experiment using human subjects on Amazon Mechanical Turk, a crowdsourcing survey platform. To study the quality of such an application, the methodology is tested in train stations, which are complex indoor environments.

First, the paper describes a device composed of a Lidar and a 360 camera that collected data from two Parisian train stations: Gare St Lazare and Gare de Lyon. It then shows how a ResNet architecture with 18 layers was trained to classify a given image into a spatial segment. By analyzing the prediction results of such identification from the output of the network, similarity is analyzed across different spatial segments within the station and a measure of legibility among indoor spaces is developed.

However, it is difficult to discern whether a DCNN model makes similar errors and uses similar visual cues as human beings do. To shed light on this problem, an online experiment is carried out to compare DCNN classified legibility and the perception of actual human subjects. In this experiment, participants are presented with an interface wherein they should select images that they believe to be spatially similar. A space recognition task on Amazon Mechanical Turk collected 4,015 samples. Results indicate that the legibility measured by the computer is highly correlated with human judgements. This is discussed further in section 6.

The application of a DCNN model to image data of spaces may be a relatively objective measurement of legibility, and statistical and quantitative analysis based on the model may be well suited to guide architectural design and management. Also, the data collection devices and pipeline can be scalably in practice.

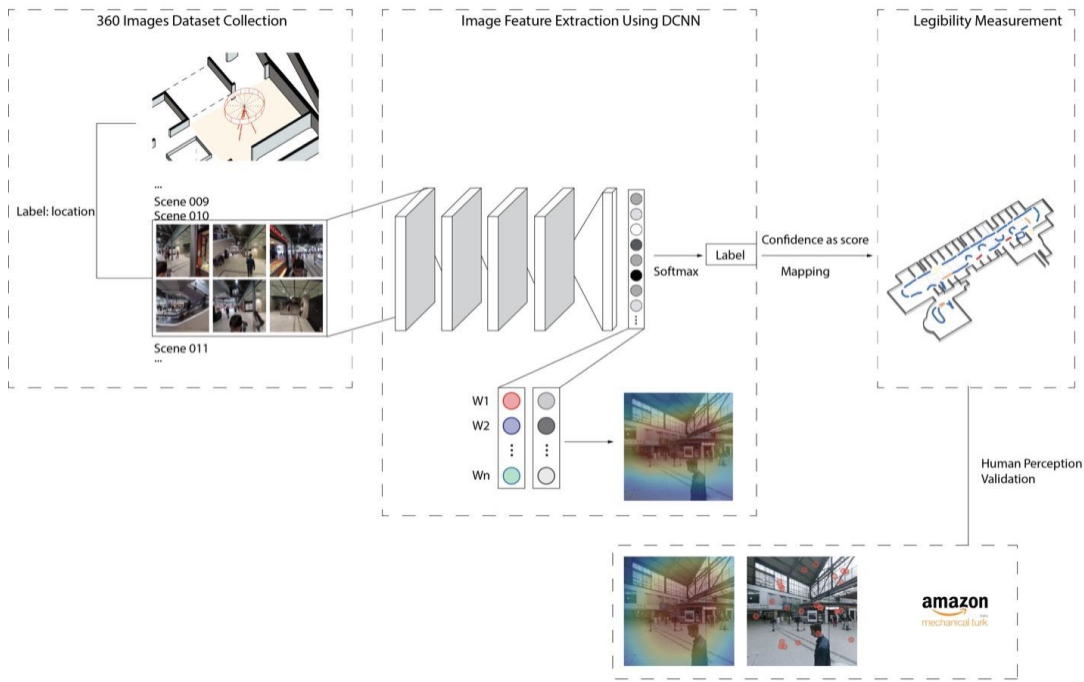

Figure 1. Diagram of methodology framework

**a. Data collecting and device design**

The study area covers most of the functional spaces (excluding tracks, boarding platforms, and administrative areas) in two train stations in Paris, Gare de Lyon and Gare St Lazare (Fig. 1). The image data are collected using a device designed by the research team for this project. (Fig. 2).

The data preparation task involved collecting as many geo-tagged images in the stations as possible. Geo-tags should be precise enough to indicate in which space the images were taken. Considering the traffic flow at stations (246,500 daily passengers at Gare de Lyon and 275,000 at Gare St. Lazare), and the desired potential to scale the application in the future, we designed a portable and parsimonious device that automates the process of taking images and documenting their corresponding coordinates. A light-weight Lidar (Velodyne Lidar Puck VLP-16) and an inertial measurement unit (IMU) to document moving trajectory, a 360 camera (Vuze Plus 3D 360 Spherical VR 4K Camera) to capture visual attributes, and a microcomputer connected to an external hard drive were installed on a refitted trolley (Fig. 2). In order to maintain the mobility of the device, it is disconnected from external electric power and a battery is integrated into the device itself. With the help of the device, Gare de Lyon and Gare St Lazare were scanned in 58 minutes and 1 hour 25 minutes respectively. The moving trajectory (Fig. 3) covers most areas of the stations open to the public—excluding platforms, the internal areas of shops, and restrooms.

**b. Data processing**

The data collected from the device consists of three components: IMU data, Lidar data, and 360° panoramic image data. First, the timestamps of those three data sources are synchronized.

Each image extracted from the camera follows spherical projection and covers a 360° view angle. In order to process the images, we transform this visual input into cubic projection and crop the panoramic images. The approximate field of view of an individual human eye (measured from the fixation point, i.e., the point at which one's gaze is directed) varies by facial anatomy, but is typically 30° superior (up, limited by the brow), 45° nasal (limited by the nose), 70° inferior (down), and 100° temporal (towards the temple) (Savino et al. 2012). In this case, panoramas were cropped into 90° each, covering almost all temporal views.

The goal of translating from spherical to cubic projection is to determine the best estimate of the color of each pixel in the final 6 cubic texture image given the spherical image. Bourke (2016) proposes a transform strategy of two stages. The first stage is to calculate the polar coordinates corresponding to each pixel in the spherical image. The second stage is to use the polar coordinates to form a vector and find which face and which pixel on that face the vector strikes.

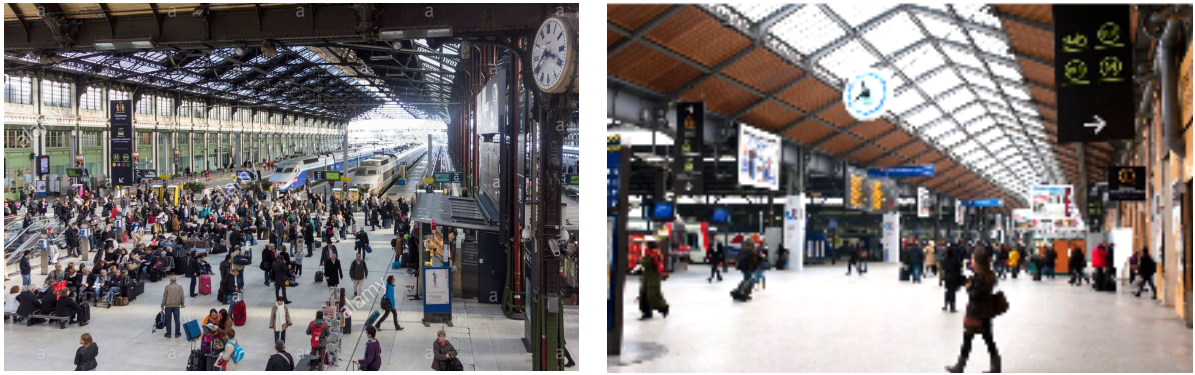

**Fig. 1** Profile View of the study area - two train stations in Paris: (a) Gare de Lyon and (b) Gare St Lazare

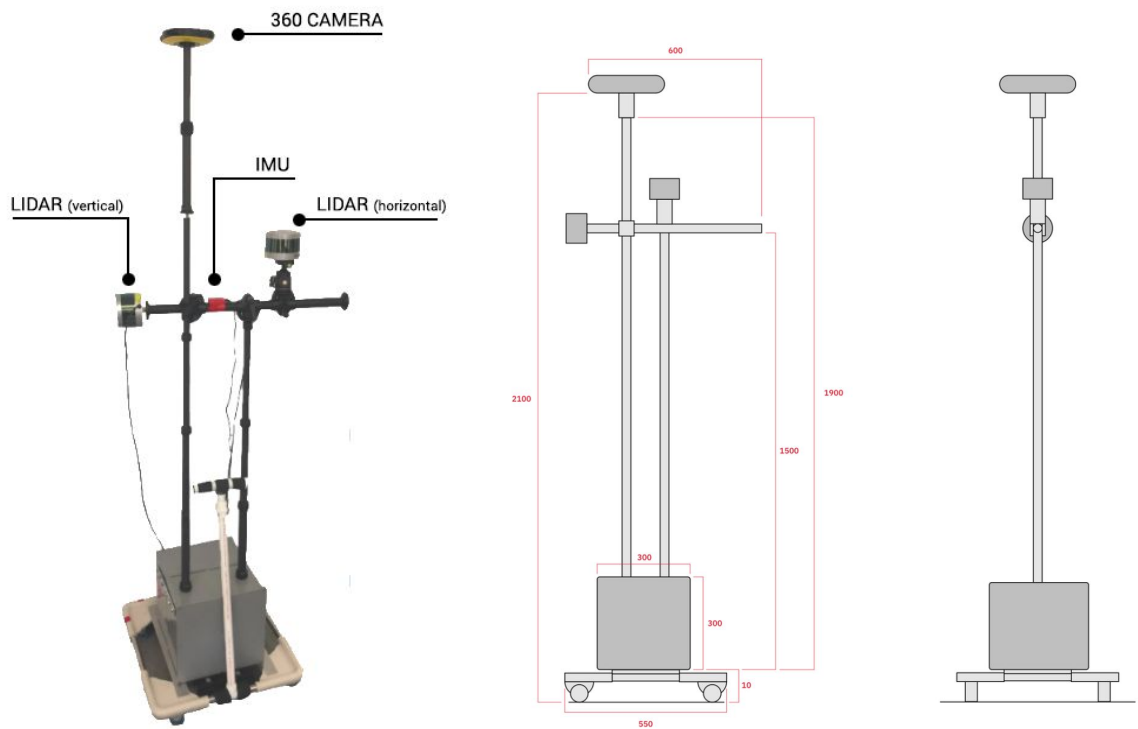

**Fig. 2** (a) An image of the device and (b) elevation drawing of the image data collection device

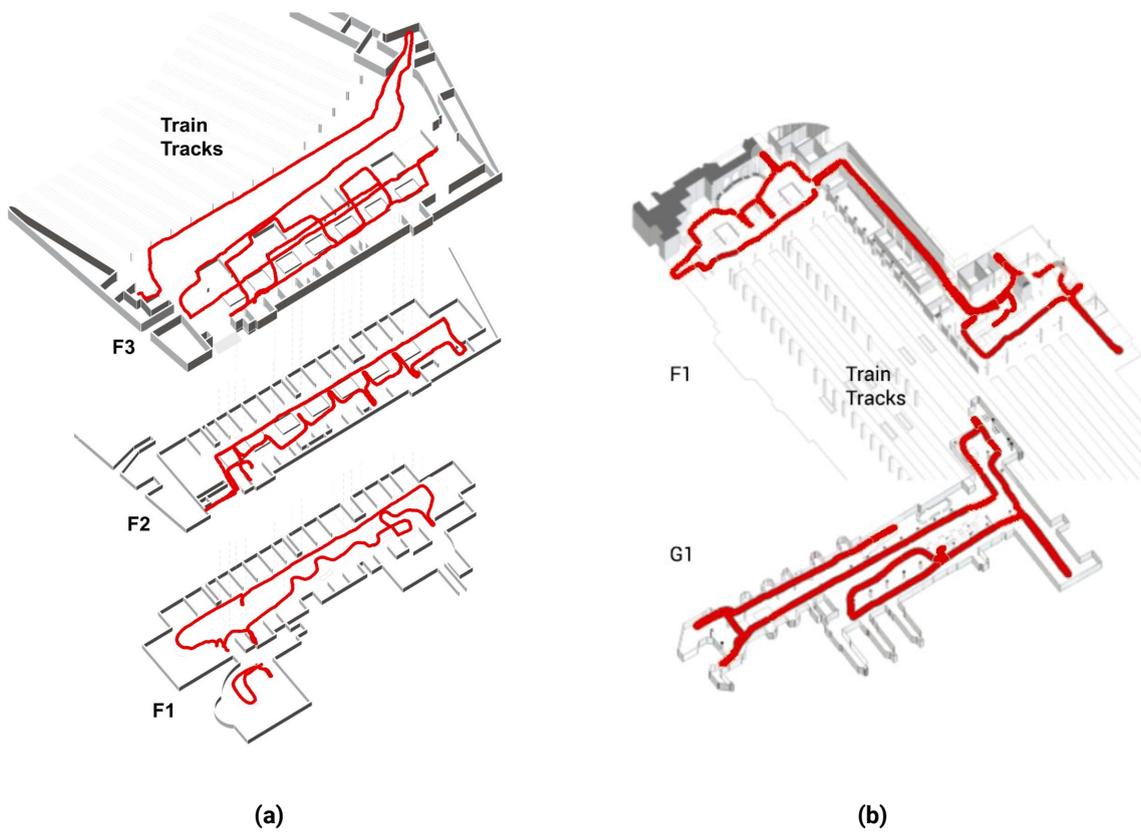

**Fig. 3** Device moving trajectory in (a) Gare St Lazare and (b) Gare de Lyon

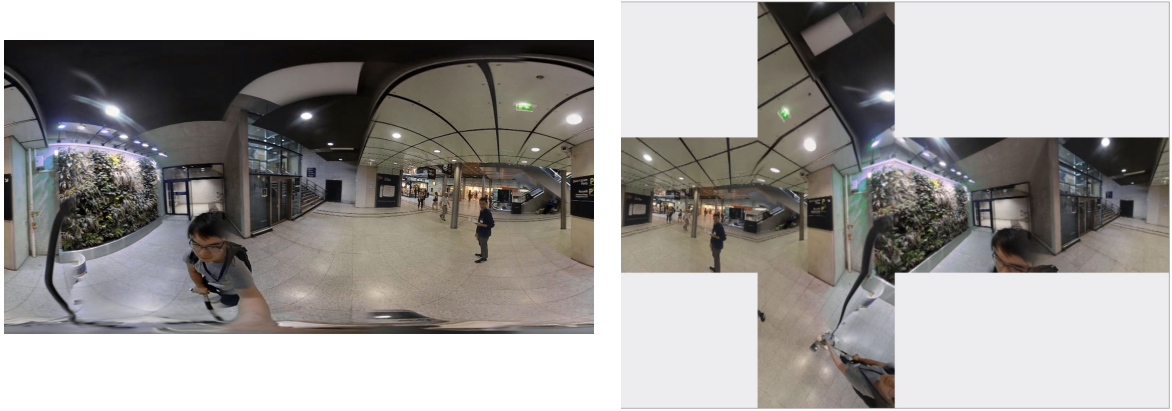

**Fig. 4** (a) Spherical projection and (b) cubic projection

The process of conversion is illustrated in (Fig. 4). Each panoramic image is cropped into 16 images, containing 8 images each rotated 45° of 0° pitch and 8 images each rotated 45° of 15 pitch.

Using open source Google Cartographer (Google, 2018) for real-time simultaneous localization and mapping (SLAM), we collected the trajectory as well as the architectural plans from the combined IMU and Lidar data. After the trajectory is acquired, each processed image is automatically annotated using its corresponding coordinates.

### c. Scene categories

The task of modeling is to identify the space, which is a classification rather than regression problem. The stations are divided into spatial segments, and each image should only be annotated with its respective spatial segment. . Three criteria are taken into consideration when designating those spatial segments: a consistent and unified architectural function (for instance, images within one waiting room should be defined as one segment); a clear boundary (wall, fence, or column); a relatively small area. Adhering to these principles, (Fig. 5.) shows an example of the location and the amount of images. Since samples size may vary at a large scale, we manually tailor the sample size to fit the area size of the spatial segment.

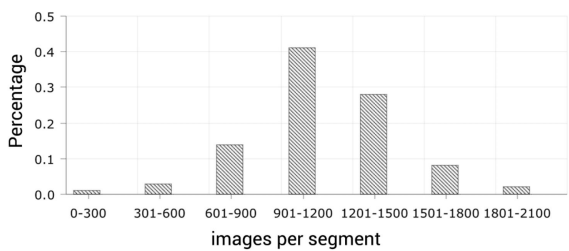

**Fig 5**. Distribution of training samples per category

### d. ResNet, the DCNN architecture

Deep Residual Network (ResNet) (He et al. 2015) (Fig. 6.) is an advanced framework to ease the training of networks that are substantially deeper than those used previously in the field. ResNet achieved state-of-the-art performance in computer vision tasks, such as object detection and scene semantic segmentation (Zhou et al. 2014) at the time of its emergence. It won first places on the tasks of ImageNet detection competition, ImageNet localization, COCO detection, and COCO segmentation. It is a very popular and typical framework and has many variations in its structure (Huang et al. 2017). In this experiment, a 16-layer ResNet is utilized.

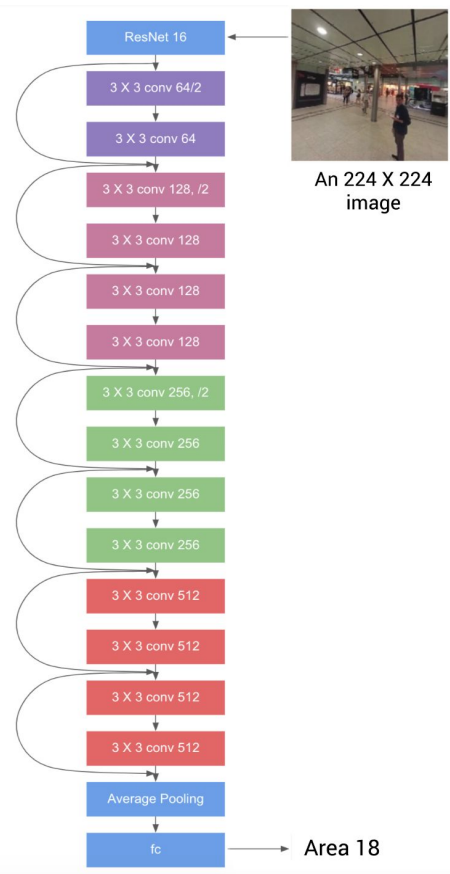

**Fig 6**. ResNet architecture

## 4. Results

### a. Data collection and data processing

58 minutes and 1 hour 25 minutes of video data were recorded in Gare de Lyon and Gare St Lazare, respectively, at 30 frames per second. We extracted panoramic images with the rate of 1 image per second from the recorded video. After the projection transformation, 50,401 and 85,993 images were cropped from panoramic images from each station. Since we found that augmentation in data-space provides a greater benefit for improving performance and reducing overfitting, in this scenario, two types of data augmentations are applied: random cropping regions, in which the area is 0.4 to 1.0 compared with the original image; and random horizontal flipping for every image.

### b. Model training and testing results

The Resnet DCNN was implemented based on Pytorch framework. It was trained using stochastic gradient descent with a constant learning rate of $10^{-4}$. The model is deployed on a workstation that consists of two parallel computing graphic card GeForce GTX 1070 Ti.

| Station | Test Accuracy Top-1 % | Test Accuracy Top-5 % |
| --- | --- | --- |
| **Gare de Lyon** | 97.1149 | 99.8897 |
| First Floor | 97.5545 | 99.9920 |
| Lower Floor | 96.8892 | 99.7563 |
| Pitch = +30 | 97.1588 | 99.8043 |
| Pitch = 0 | 97.8109 | 99.9720 |
| Pitch = -30 | 96.3437 | 99.8323 |
| **Gare St Lazare** | 97.2317 | 99.9904 |
| Third Floor | 97.2231 | 99.9859 |
| Second Floor | 98.9124 | 1.0000 |
| First Floor | 96.5612 | 99.6920 |
| Pitch = +30 | 97.0101 | 99.9189 |
| Pitch = 0 | 97.9731 | 99.9879 |
| Pitch = -30 | 96.4228 | 99.8210 |

**Table. 1** Overall testing accuracy

For Gare de Lyon, we tested the model on 88,869 images with similar spatial distribution with control images, and 86,305 images were correctly predicted for their spatial segments. The model achieves 98.66% prediction accuracy overall (Table.1) of its top-1 choice in localizing these testing images. The first floor, mainly comprised of waiting areas and shops, achieves 97.55%, whereas the lower floor, primarily used for connection between modes of transport, has 96.89% accuracy. Meanwhile, there are some differences between horizontal perspectives (pitch = 0) and non-horizontal (pitch ≠ 0) perspectives. The non-horizontal perspectives, which show larger percentages of floor and ceiling (-30 and +30, respectively), have lower accuracy. This is probably due to their homogeneity regarding geometric patterns and materials, with few distinctive features. We also visualized confidence of different spatial segment using gradient colors (Fig. 5). Image examples of low and high confidence is shown in (Fig. 6)

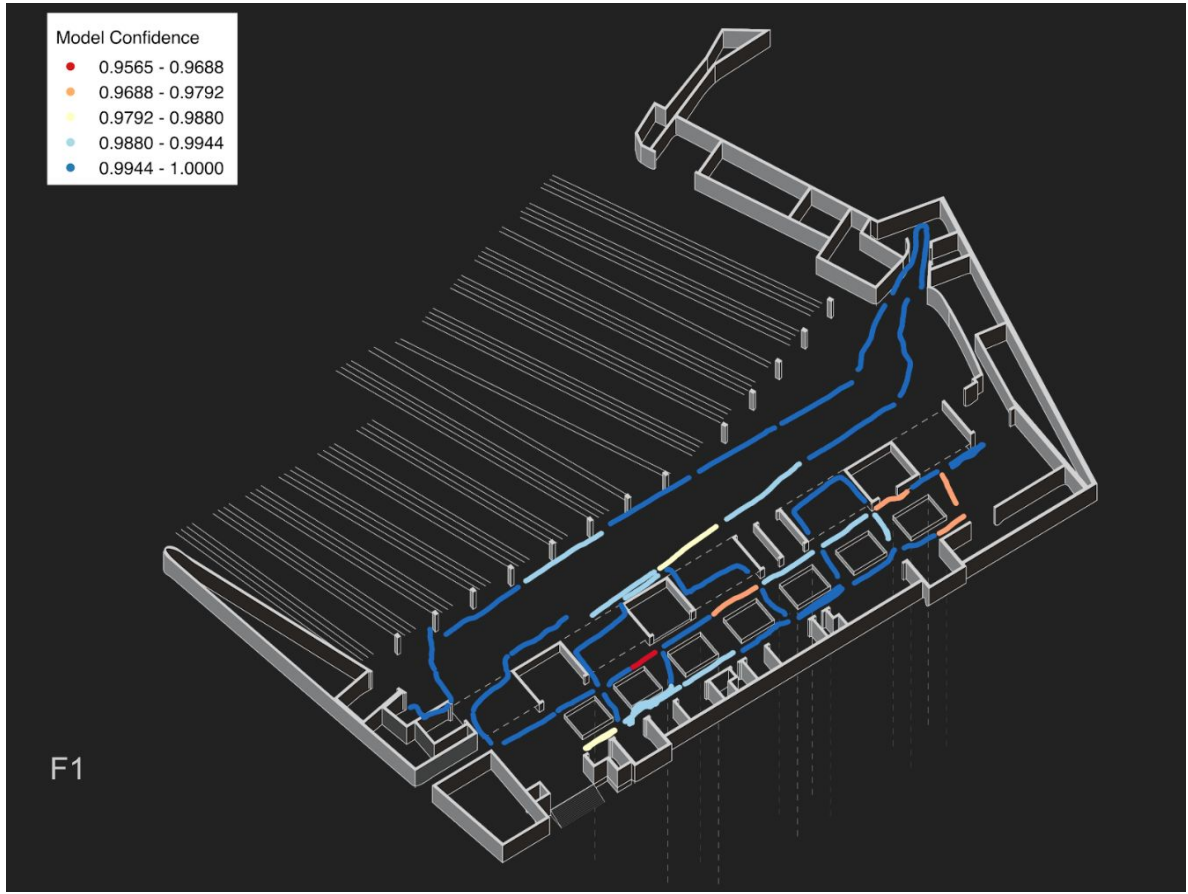

**Fig. 5** Model accuracy in each spatial segment in 1st floor in Gare St Lazare

(a) low confidence images

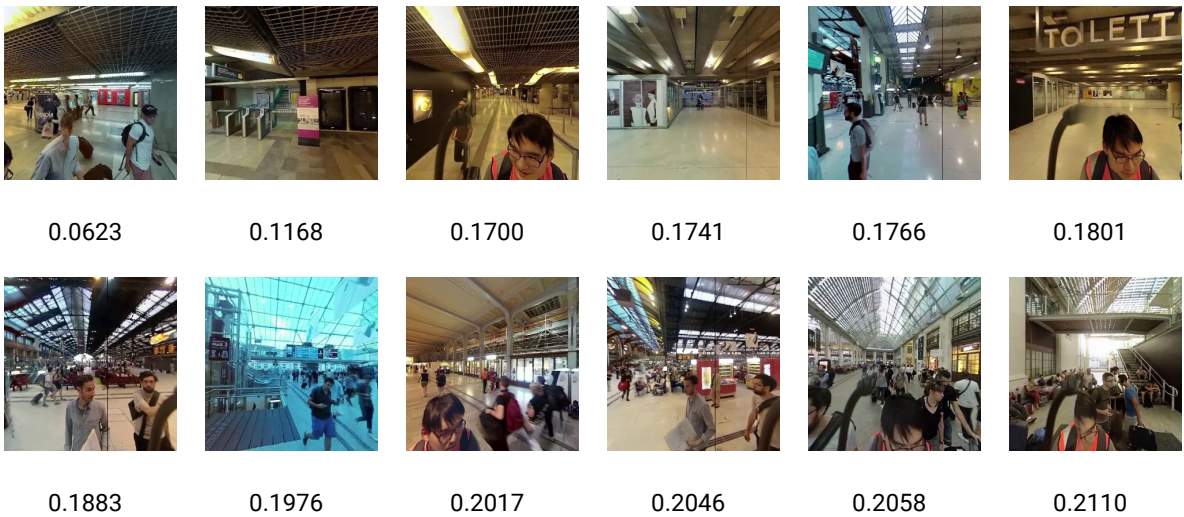

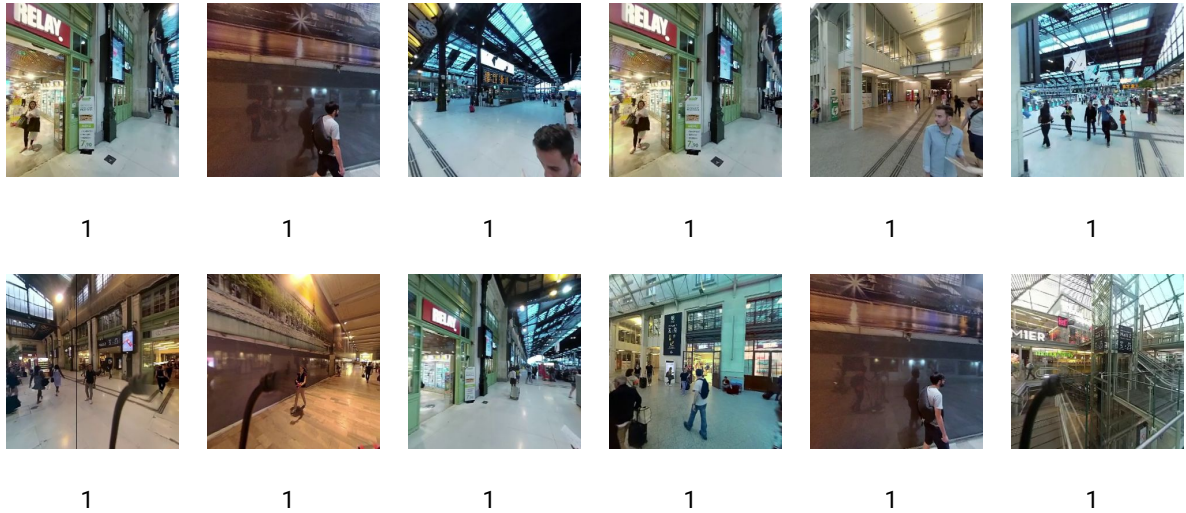

(b) high confidence images

**Fig. 6** Images with (a) high and (b) low confidence in the prediction in Gare de Lyon

## 5. Analysis

The spatial segment prediction accuracy reflects whether a scene can be correctly located into its original space while the confidence of the model can be interpreted as the degree of confidence in localizing such scene image extracted from a space. At this point, we assume that the legibility of certain spatial segments can be reflected through the DCNN model confidence. The legibility of different architectural programs as well as the visual elements that determine legibility can be quantitatively studied by analyzing the results of the model.

### a. Architectural programs

As an architectural concept, "program" is defined as the activities and functions of a particular space. Generally speaking, distinct architectural programs have a different and appropriate legibility score because their user experience is intrinsically tied to their function. For instance, the circulation and directionality of malls may not be quite straightforward in order to extend the time spent within the space. In other spaces, such as corridors in stations, clear indicators shorten wayfinding time and alleviate crowdedness and thus may be more prominent. Specific to this study of train stations, waiting areas are a key typology of space.

As such, each spatial element is tagged as one of three basic categories within the train stations: either commercial, waiting or corridor area. After averaging the confidence by programs listed above, St. Lazare has much higher legibility level in waiting areas, albeit with a relatively small sample compared to Gare de Lyon (Table. 2). Commercial areas in St Lazare are likely to cause confusion, according to the model. On the contrary, results from Gare de Lyon shows that commercial areas are highly distinguishable, while corridors are more visually confusing (Table. 2).

| Type | COMMERCIAL | | WAITING | | CORRIDOR | |
|---|---|---|---|---|---|---|
| Station | Lyon | St Lazare | Lyon | St Lazare | Lyon | St Lazare |
| Total Image | 23711 | 45287 | 34280 | 21328 | 30878 | 6420 |
| Confidence | 98.6% | 95.9% | 97.6% | 99.3% | 92.3% | 95.6% |

**Table. 2** Confidence by architectural function

### b. Building style and building age

Preliminary results showed that legibility is related to architectural style and building age. Gare de Lyon has undergone two major changes after its construction: Hall 1 has retained its original spatial qualities since the 1900s; Hall 2 was renovated in 2012; Hall 3 went through an expansion project in 1977. The model performed best in Hall 1 and worst in Hall 3 (Table. 3). The design of Hall 1 consists of a combination of materials including iron, plate glass, colored tile and reinforced concrete. The contrast between the brightly colored panels and the vertical columns is strong. Those architectural characteristics helped the model to recognize space. However, Hall 3, with bland modern materials and repeated architecture elements, is hard for the model to identify.

| Hall | Hall 1 (1900s) | Hall 2 (2012) | Hall 3 (1977) |
|---|---|---|---|
| Accuracy | 99.1 | 92.1 | 82.9 |

**Table 3**. Model accuracy aggregated by Halls

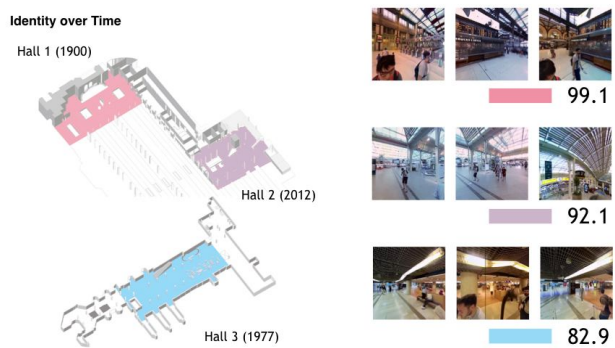

**Fig. 8** Model accuracy aggregated by Halls

### c. Discriminative regions and repetitive elements

Certain elements in a physical environment may significantly increase the legibility level. Therefore, it is a necessary task to find and locate those elements which improve people's wayfinding performance. Drawing from a number of recent research projects in deep learning, it is possible to infer the implicit attention of CNNs on an image by simply looking at the weight matrix.

Recent works in DCNN model feature have shown that the model retains a remarkable ability to localize objects in the convolutional layers until the final fully-connected layer, which increases the potential to identify the discriminative image region in a single forward pass process. Classification Activation Mapping (Zhou et at., 2016) is a decent way to implement this idea. Combining the Resnet18 model architecture, we describe the detailed procedure in terms of CAM technique: in a forward pass, we define $F_k(x,y)$ as the last convolutional layer feature map $k$, and define $\omega$ as the weight vector between global average pooling layer and softmax layer. According to the final classification result, we will have the best matching class c and its corresponding weight $\omega_c$. Here we take ωc as the best linear combination weights for feature map $F_k(x,y)$, in terms of obtaining the importance of the activation at feature map space. Hence, the class activation map can be given by:

$$M_c = \sum_k \omega_c F_k(x, y)$$

The size of Mc would be the same as the last convolutional layer, in our case, 11 by 11. We only need to upsample it to 224 by 224, and composite the original input image. It is possible to identify the images' regions most relevant to the class, to further understand the visual spatial feature that the model is most interested in.

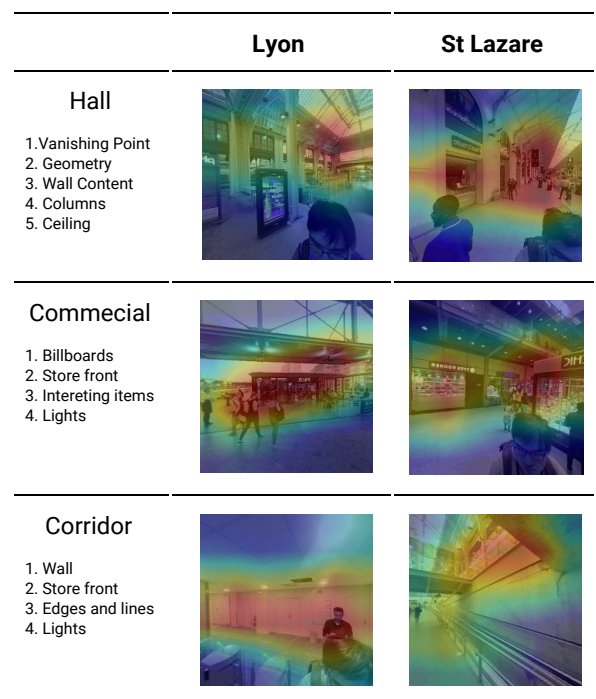

**Fig. 9** Examples of heatmaps of two station

By observing the heatmaps (Fig. 9), it is possible to recognize and understand the decision process of the model. In images with more details and information, such as the halls of both stations, the models tends to rely upon items with higher level features and with semantic meanings, while in simpler spaces such as staircase or corridors, the model is looking at low-level features such as edges or lines on the walls, or textures on ceiling and floor. Notably, when images have an obvious vanishing pointing, the model would prioritize its attention on the vanishing point instead of other strong elements. In other scenarios, model would prioritize detailed textures over raw textures, curved geometry over orthogonal geometry. That is to say, model has an attention order from perspective, object, texture to geometry (from high to low) (Fig. 10).

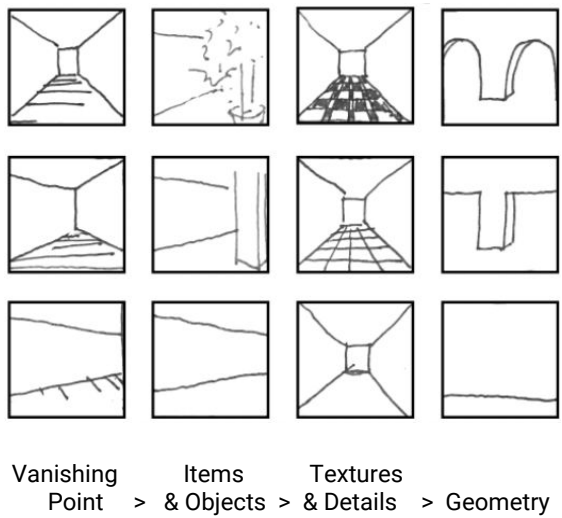

Vanishing Point > Items & Objects > Textures & Details > Geometry

**Fig. 10** Model attention priority(Author's diagram)

d. **Spatial correlation and spatial similarity structure**

In order to figure out the similarity between two spatial segments, a covariance matrix is introduced. Let *i* denote the sequence of each spatial segment, *j* and *k* denote two spatial segments, N as sample total count:

$$q_{jk} = \frac{1}{N-1} \sum_{i=1}^{N}(X_{ij} - \overline{X}_j)(X_{ik} - \overline{X}_k)$$

In this equation, substitute $X_i$ using the input before the last softmax layer of DCNN architecture. In St. Lazare's case, $X_i$ is a 111 by 1 vector.

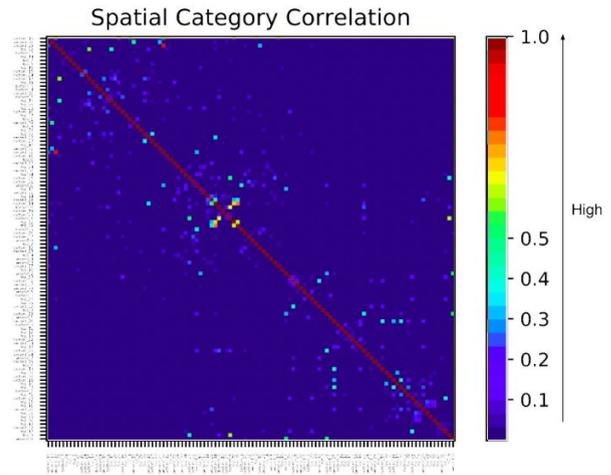

**Fig. 11** Correlation between spatial segments in Gare St Lazare

From the graph, it could be interpreted that most spatial segments are not very similar to each other. However, there are certain areas where two segments look visually very alike. To further investigate where and why those similarities occur, it is necessary to map it spatially.

Room-level 3D models for each station have been built to better visualize the correlation in a spatial context (Fig. 11). The width of lines between two spatial segments represents the similarity (thicker representing more similar). In Gare St. Lazare, two obvious characteristics have appeared: for Gare St. Lazare, links between spaces in one side of the corridor are strong while links across opposites corridors are significantly weaker; for Gare de Lyon, there are almost no links between the two floors and all stronger links are within a single floor. That is to say, Gare St. Lazare has a vertical visual layer structure and Gare de Lyon has a horizontal visual layer structure.

e. **Clustering**

To find clusters in a number of spatial segments, we used the Louvain Method, which extracts communities from large networks. (Blondel et al., 2008). The method is a greedy optimization method that appears to run in time $O_n$. Let $A_{ij}$, $k_i$, $k_j$, $2m$, $c_i$, $c_j$, $\delta$ represent: the edge weight between nodes *i* and *j*; the sum of the weights of the edges attached to nodes *i*; the sum of the weights of the edges attached to nodes *j*; the sum of all of the edge weights in the graph; the communities of the nodes *i*; the communities of the nodes *j*; simple delta function. The modularity can be defined as:

$$Q = \frac{1}{2m} \sum_{ij}[A_{ij} - \frac{k_i k_j}{2m}\delta(c_i, c_j)]$$

This value is calculated by two steps: (1) removing $i$ from its original community, and (2) inserting $i$ to the community of $j$. The two equations are quite similar, and the equation for step (2) is:

$$\Delta Q = [\frac{\Sigma_{in} + 2k_{i,in}}{2m} - (\frac{\Sigma_{tot} + k_i}{2m})^2]$$
$$- [\frac{\Sigma_{in}}{2m} - (\frac{\Sigma_{tot}}{2m})^2 - (\frac{k_i}{2m})^2]$$

The diagram (Fig. 12) reflects visual similar spatial segment patterns. The closer the points, the more similar they are. In St Lazare, for instance, although most points keep certain distance with one another, there are a region where 10 points almost overlapping with each other in the diagram. These 10 spatial segments are commercial spaces with similar visual characteristics. The clustering analysis can be applied in assisting spatial management by discovering visual clusters to reduce confusion between spaces.

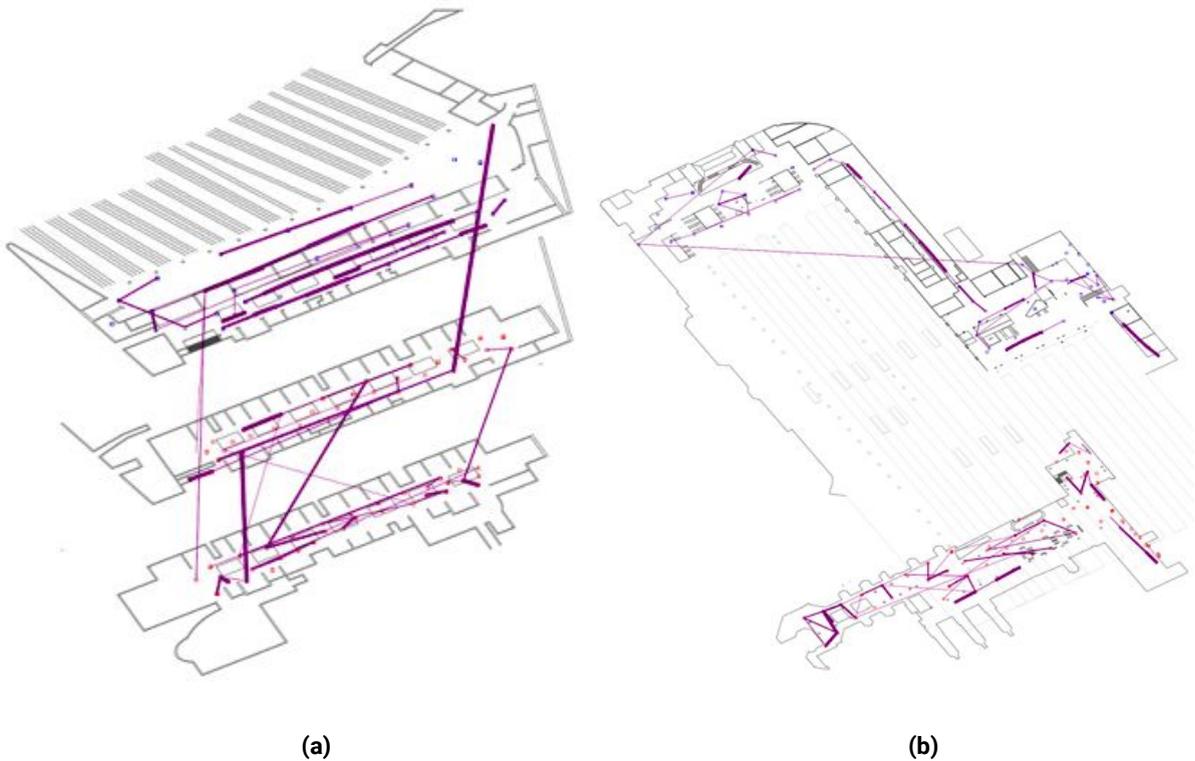

(a)            (b)
**Fig. 13** Visual similarity structures in (a) Gare St. Lazare (b) Gare de Lyon

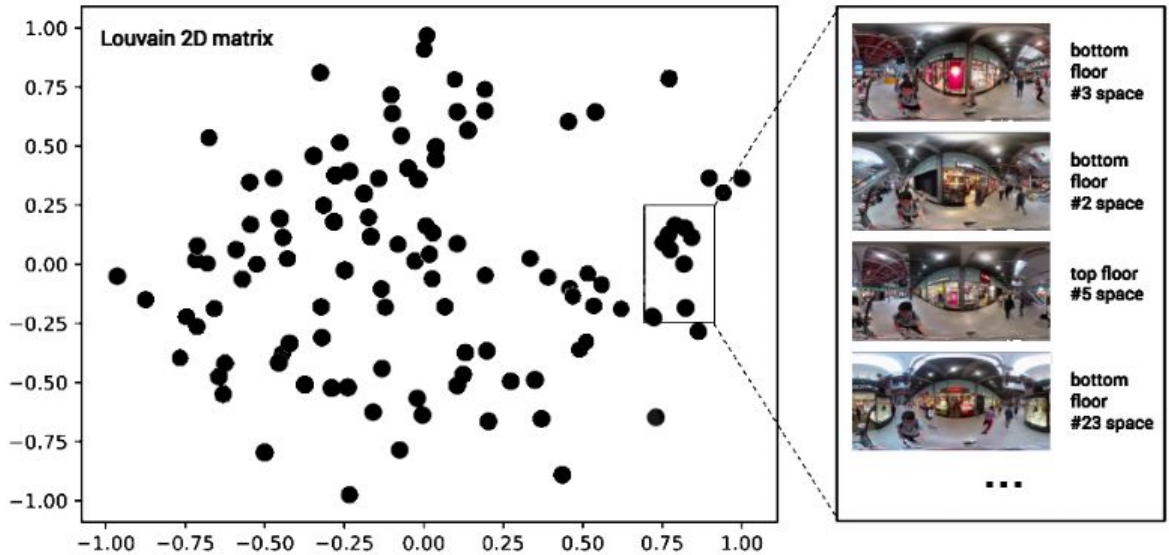

**Fig. 14** Louvain 2D Matrix for Gare St. Lazare

## 6. Validation

### a. Introduction of validation

Computer vision methods that quantify the perception of the built environment are a promising methodology compared to previous experiments utilizing interviewing methods, which are heavily limited by the sampling numbers, time, and cost. In order to verify that computer vision can be effectively used to study the relationship between a physical space's appearance and human perception, a validation survey is implemented. The survey is designed to 1. compare the results of legibility evaluation predicted by CNN model with perceptions evaluated by humans; and 2. compare the visual clues that support the CNN model with those humans used to make decision.

More recently, online data collection methods where humans evaluate images or crowdsourcing (P. Salesses, 2013) have increased the ability to externalize individual's perception. By using online visual surveys, the availability of participants has also exponentially increased. It enables us to control for similar conditions within the comparison.

Thus, a survey website (Fig. 15), named SpaceMatch, implementing interactive features with easy-to-follow instructions was launched on Amazon Mechanical Turks to enable participants to answer required questions:

### b. SpaceMatch

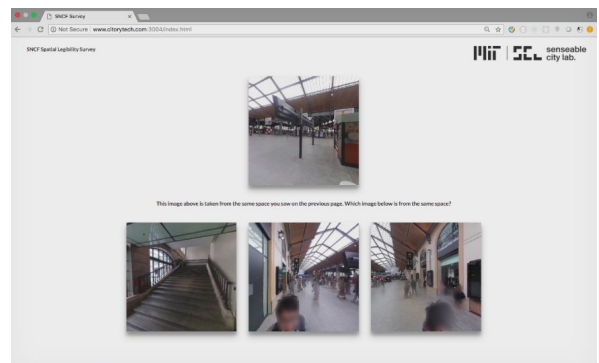

**Fig. 15** User Interface of SpaceMatch

SpaceMatch takes advantage of web animation to maximize similar conditions between the human and CNN model comparison. In each section, participants are given an animated rotating panoramic pictures providing information of one chosen spatial segment A. The animation lasts 10s so the participant can capture details of the scene. On the next page, one image snippet from the spatial segment A is displayed as control image with a legend explaining that it is taken from the scene they just saw. It is then followed by three image snippets:

**IMAGE 1:** randomly picked from the same spatial segment A,
**IMAGE 2**: randomly picked from spatial segment B that follows the maximum covariance principle

(explained in next section) computed by the CNN model

**IMAGE 3**: randomly picked from one spatial segment C excluding A and B

The three images are shuffled before displayed on the page. The participants are then asked "Which image of the three is from the same space (as the one above)".

After the participants decide on the image from the same space, they are asked to point out three features that help them to make the decision. Every time they click on certain area, the area is marked and the participants are asked to choose one property from a drop down list, including light, material, texture, geometry or object. The control image is displayed on top of the page as references.

Each participant is asked to complete five questionnaires. All the results are stored in the database with a set of attributes (Table. 4).

| | |
|---|---|
| User information | IP of the participants to avoid robotics |
| Image ID of A,B,C | Filename to orient spatial segments of the displayed images on the webpage |
| User Choice | Filename and its spatial segments of the image chosen by the user |
| Location of pointing(*3) | X,Y coordinate of the click position on the image |
| Properties of pointing(*3) | Properties of light, material, texture, geometry, object that user chose |

**Table. 4** Attributes associated with each question

### c. Sampling

The purpose of the questionnaire is to verify if the CNN model's evaluation results of the physical space can reflect human perception. The question is then translated to test if highly similar spatial segments defined by the CNN model are also highly similar in human's perception. So the maximum covariant segments pair is selected as sample principle in the survey:

After the last activation layer, the dense (fully connected) layers of the CNN model performs classification on the features extracted with a softmax activation function to generate a value between 0-1 for each spatial segments. The vector $y_i$ consists of value of each spatial segment. We define $P_{ij} = \frac{1}{K}\sum_{k=1}^{K} y_j^i + \frac{1}{D}\sum_{d=1}^{D} y_i^j$ as the covariance between $(i,j)$ segments, where $y_j^i$ represents the predicted value of $j$ segment at $k$ sample belonging to $i$ segment, $D, K$ are the total sample amounts of $(i,j)$ segments.

For each sample from segments, the probability of prediction among segments is:

$$y_i^j = \frac{e^{-z_i}}{\sum_{n=1}^{N} e^{-z_n}}$$

For each segment $i$, the maximum covariance with segments $j$, $\widehat{P}_{ij}$ represents that segment $j$ is highest similar compared to other segments in the model in our case with maximum probability.

All the segment covariance $\widehat{P}_{ij}$, $i \in \{segment1, segment2, ...\}$, $j \in \{segment1, segment2, ...\}$ are ranked and images from the top 90 pairs are selected as question pool. In the dataset of the question pool, the control image from $i$ segment is assigned to ID: image_a_0. One image picked up from the same $i$ segment(class) is assigned to ID: image_a_1. One image from $j$ segment which has the highest value of $\widehat{P}_{ij}$ is assigned to ID: image_b. In other word, Image_b as explained above is from the segment regarded as the most similar segment by the model. Meanwhile, a random picture from neither $i$ segment nor $j$ segment is assigned to ID: image_c.

### d. Validation results

The survey was launched on Amazon Mechanical Turk for 21 days and 4015 valid results (after ruling out survey robotics) (Table. 5) were collected in total. 54.5% of the participants choose the correct segment and 37% participants choose image_b, which represents the most similar segment defined in the model, but not include in the sample video. Only 8.5% of the participants selected image_c (which differs from the sample video, according to the DCNN results) showing that they had a different judgement compared to the model. This result means that the spaces that confused the CNN model have a very high probability to confuse humans as well. The results from the crowdsourcing survey can validate that quantifying the visual legibility of built-environment by computer vision is sufficient to represent human's confusion.

| Image ID | image_a_1 (same segment) | image_b (highly similar with defined by model) | image_c (random segment) |
| --- | --- | --- | --- |
| selection (times) | 2187 | 1484 | 344 |
| % | 54.5% | 37.0% | 8.5% |

Table. 5 Total selection amount among image categories

### e. Feature comparison

Even though the result of the CNN model for visual similarity earns similar result to human visual perception in the indoor space, a further study is processed to compare which visual features human and the CNN model used in their decision-making process.

As mentioned before, the Class Activation Mapping (CAM) method is applied after the fully-connected layer of the model to visualize activation features in the CNN model (Zhou et al., 2016). For the human perception, in order to understand how the participants make their decision when choosing the images of the same space, the survey asks them to pin three points. These points are also visualized on the images and then transformed into a heatmap representation to compare with the heatmap generated from computer vision.

Images below come from eyeballing comparing between the heatmap generated by the CNN model and human's selection. In some spaces, they share quite similar focusing area. While in some spaces, human and the CNN model "look" at different areas.

(c) spaces model and human share similar features

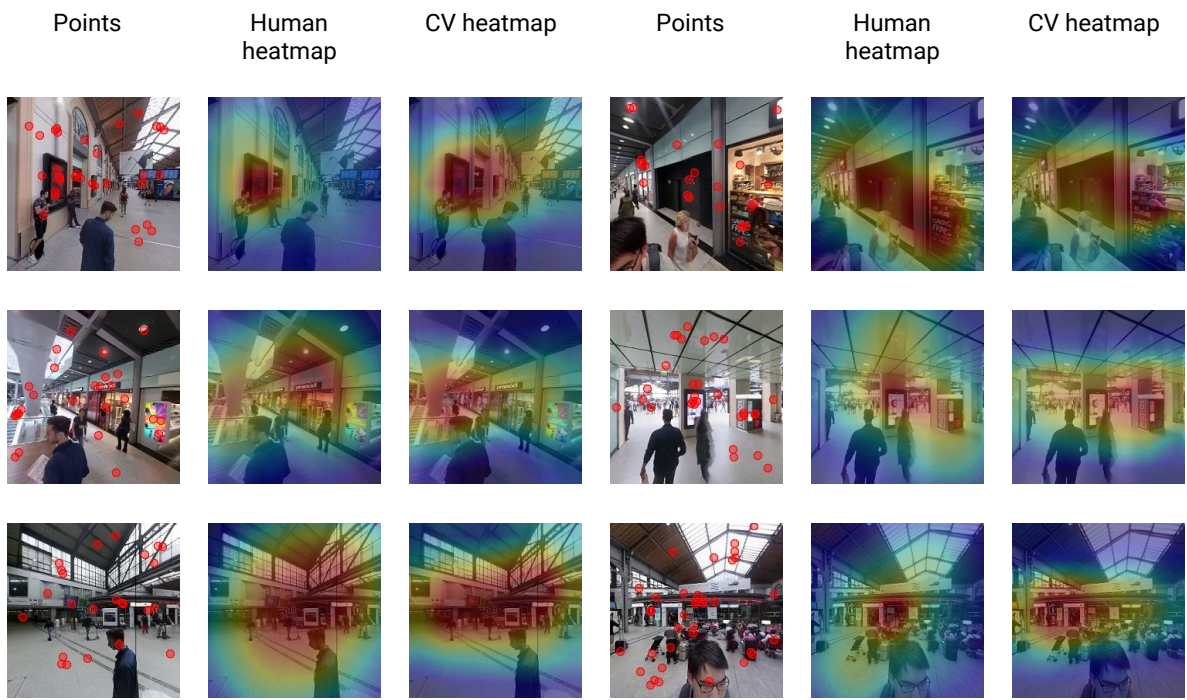

| Points | Human heatmap | CV heatmap | Points | Human heatmap | CV heatmap |

(d) spaces model and human look at different features

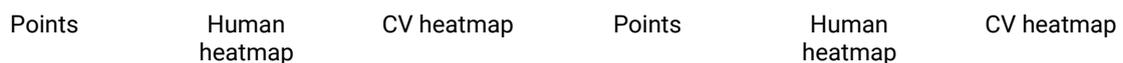

| Points | Human heatmap | CV heatmap | Points | Human heatmap | CV heatmap |

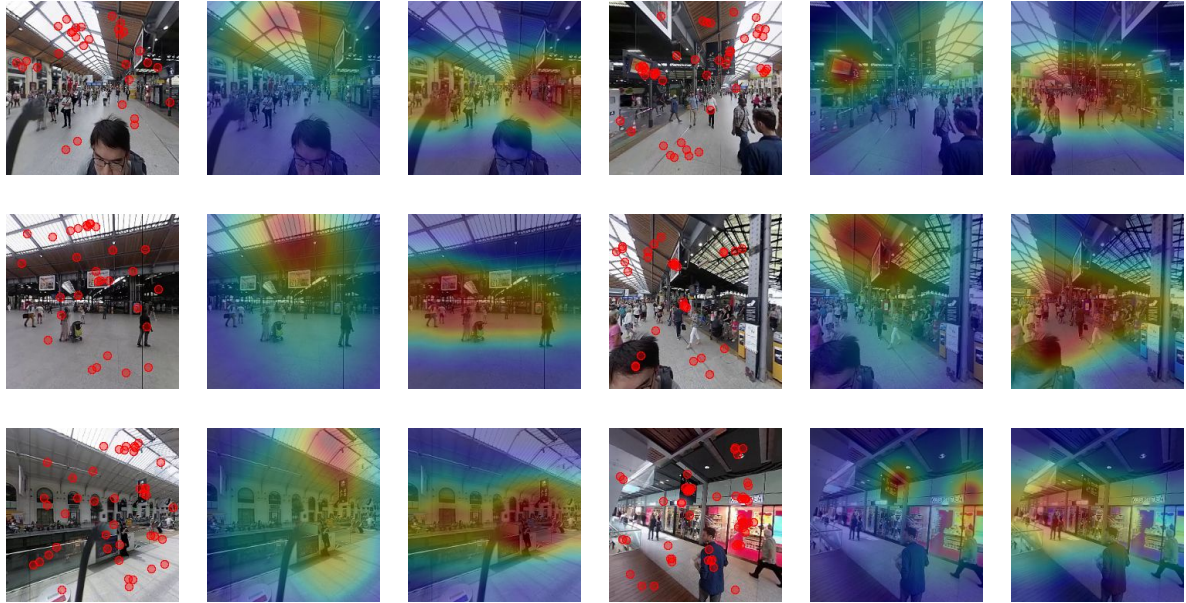

**Fig. 16** Images of (c) similar and (d) different attention area between computer model and human decisions

Compared to the focus features of CNN model, human's spatial legibility is heavily dependent on objects. The selected properties and points show evidence that the participants used spatial features such as TV screens or advertisement boards to help recognize spaces and locate themselves. Next, they tend to use materials to help position themselves: for example, the consistent texture on the floor or on the roof always refers to one consistent space. The different feature can also be explained that human always unconsciously transform the input information of a image to a 3D reconstructed spaces. However, for the CNN model, even though it is trained by panoramic images, it can only extract shared 2D information. It lacks human's "spatial feeling". We summarize human attention priority in **Fig.17**.

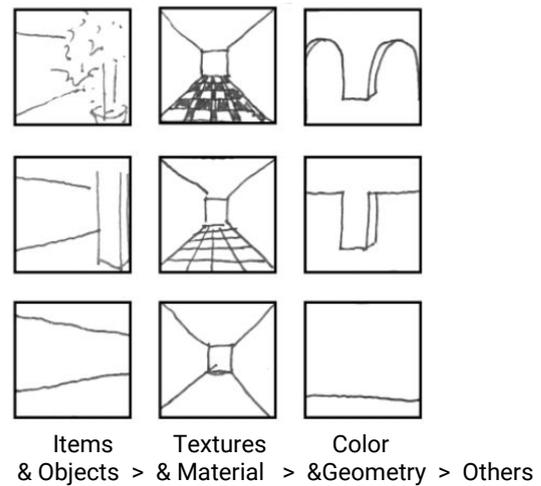

Items & Objects > Textures & Material > Color &Geometry > Others

**Fig. 17** Human attention priority(Author's diagram)

**f. Quantitative comparison**

To further quantify the overlapping features between human and model, we compared the focus areas picked by participants and the activation heatmap of the neural network model by posterior predictive check. The focus areas, sampled by a 10-pixel radius circle area, are mapped onto the gray heatmap. As explained above, the value from the global average pooling outputs represents the spatial average of the feature map of each unit at the last convolutional layer generated by the model. The posterior predictive check here is to quantify to what extent participants' selections follow in model's confidence interval conditioned on the maximum probability they can have：

| Properties | 1st Choice | 2nd Choice | 3rd Choice |
|---|---|---|---|
| Object | 41.5% | 33.8% | 30.5% |
| Material | 15.3% | 20.0% | 20.7% |
| Color | 13.9% | 15.6% | 16.5% |
| Light | 15.5% | 16.6% | 13.0% |
| Geometry | 10.5% | 10.4% | 12.7% |
| Other | 3.3% | 3.6% | 6.7% |

**Table. 6** Total selection amount in properties

$$\eta = \frac{\sum\sum_{x \in A, y \in A} P(x,y)}{\sum\sum_{x \in S, y \in S} \hat{P}(x,y)}$$

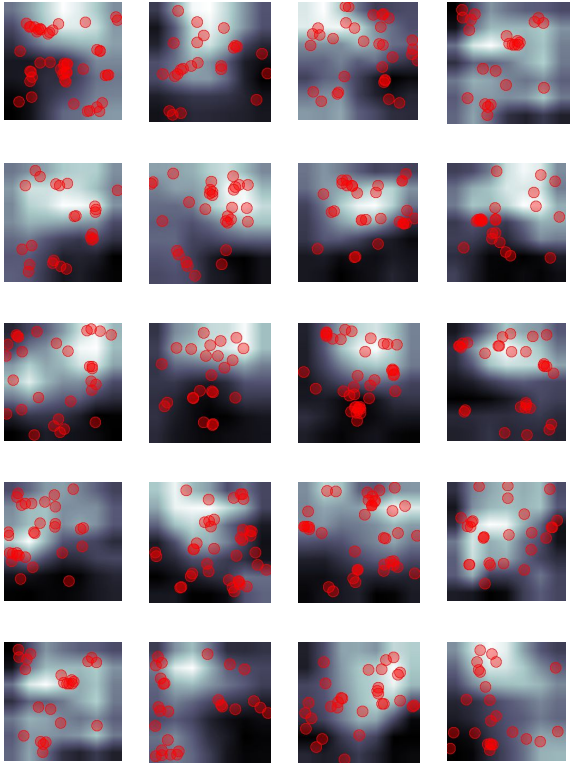

**Fig. 18** Posterior predictive check by mapping points on one channel heatmap

| η Value | Max | Min | Mean | Medium |
|---|---|---|---|---|
|  | 0.756 | 0.202 | 0.465 | 0.491 |

**Table.7** Distribution of the posterior predictive check

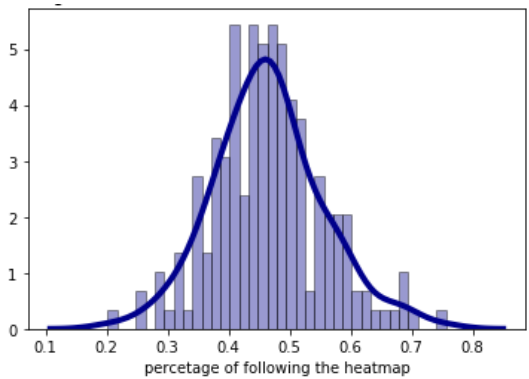

**Fig. 19** Histogram of posterior predictive check of human selection

Among all the survey images, the maximum value is 0.756 which refers to that in some scenes, a human's focus area highly follows, to the extent of 75.6 percent, the distribution of the most informatic area in the model. However, in some other scenes for the minimum condition with a value of 0.202, humans use only 20.2 percent informatic area in the model as reference for localizing themselves and look mostly at totally distinguished area compared with the model.

Based on the above analysis, the questionnaire shows that it is clear that the perception results (to match the correct spatial segments) by model and human are highly similar to each other. The model and the humans are confused by the same spatial area, and make the same wrong decisions. However, in most of the cases, they choose different features to make such decision: the CNN model heavily relies on the low resolution information in the image while for human, the perception process is much more complex.

## 7. Discussions

In this paper, we introduce a complete workflow to quantify DCNN models on a large-scale dataset collected in indoor spaces. The model performed very well in recognizing spaces. The preliminary results reflect differences in legibility in different spaces in the station. Also, we developed a methodology to validate using survey results from a crowdsourcing platform. The correlation between the model result and survey result confirmed the validity of our legibility modelling proposal. The current study suggests the following contribution from our work:

- Our self-designed device enables us to collect imagery data with geo-information in very short time. Using consumer level Lidar, IMU and 360 camera, our prototype can be easily reproduced for quantifying legibility or similar tasks in indoor and outdoor spaces alike.

- With data augmentation and a training parameter, our model has achieved 98% top-1 accuracy on testing set. This technique can be further applied to indoor positioning by virtue of its parsimony and deployment speed.

- The analysis of the model's accuracy on different spatial segments provides insights about quantitative discrepancy in legibility for different spaces. Along with basic architectural properties, the knowledge on legibility can further be transformed into tools for

- architectural study and criterias for interior design.

- Our design in surveying people's perception of legibility using crowdsourcing platform provides a novel strategy for conducting surveys on human perception. First, the utilization of 360 images and videos makes experiment more similar to experiences in real-time scenario. Further, the ability to scale the survey on the platform reduces individual biases.

Although this method has clearer value proposition and more reliable workflow compared to other recent works. There are still two limitations:

- Although image representations are very comprehensive, they only contain visual information. Legibility is connected with multiple senses, such asi hearing, smell and touch (Lynch, 1960). Our future research will be focusing on the justification of the argument that visual sense in the major factor in perceiving legibility.

- To what extent individual factors vary from universal perception of legibility, is still to be substantiated by more evidence. In our research, we ignored the individual factors in order to arrive at a more general and universal conclusion. In future research, through analyzing user portrait through data collected from online survey platform, we hope to find more answers to this question.

## 8. Acknowledgement

The authors would like to thank Dover Corporation, Teck, Lab Campus, Cisco, SNCF Gares & Connexions, Brose, Allianz, UBER, Austrian Institute of Technology, Fraunhofer Institute, Kuwait-MIT Center for Natural Resources, SMART-Singapore MIT Alliance for Research and Technology, AMS Institute Amsterdam, Victoria State Government, and all the members of the MIT Senseable City Lab Consortium for supporting this research. We thank our colleagues who provided insight and expertise that greatly assisted the research, in particular Stefan Seer (AIT), and Etienne Burdet (SNCF).